\documentclass[10pt,a4paper,twocolumn]{article}
\usepackage[english]{babel} 
\usepackage[top=2.8cm,bottom=2.8cm,left=2.cm,right=2.cm]{geometry}  
\usepackage{titlesec}
\usepackage{color,soul}
\usepackage[utf8]{inputenc} 
\usepackage[affil-it]{authblk}
\usepackage{amsmath,amsfonts,amssymb}
\usepackage{varioref}
\usepackage[colorinlistoftodos]{todonotes}
\usepackage[pdftex]{hyperref} 
\let\OLDthebibliography\thebibliography
\renewcommand\thebibliography[1]{
  \OLDthebibliography{#1}
  \setlength{\parskip}{1pt}
  \setlength{\itemsep}{1pt plus 0.3ex}
}
\usepackage[binary-units = true]{siunitx}
\usepackage{float}
\usepackage{tikz}
\usepackage{xcolor}
\usepackage{comment}
\usepackage{braket}
\usepackage{nicefrac}
\usepackage{caption}
\usepackage{graphicx}  
\graphicspath{{./Figures/}{./PSTricks/}}  
\usepackage{multirow, makecell}
\usepackage{appendix}
\usepackage{setspace}
\usepackage{cite}
\usepackage{csquotes}
\usepackage{array,booktabs}

\usepackage{enumitem}
\usepackage{caption3} 
\DeclareCaptionOption{parskip}[]{}
\usepackage[font=small,labelfont=bf]{caption}
\usepackage[final]{pdfpages}
\usepackage[colorinlistoftodos]{todonotes}
\usepackage{tikz}
\usepackage{abstract}

\usepackage[export]{adjustbox}
\usepackage{dblfloatfix}
\begin{document}
%\maketitle
\title{\textbf{Characterization and stability measurement of deployed multicore fibers for quantum applications}}
\author[1,*]{\small Davide Bacco}
\author[2,3]{\small Nicola Biagi}
\author[2,4]{\small Ilaria Vagniluca}
\author[5]{\small Tetsuya Hayashi}
\author[6,7]{\small Antonio Mecozzi}
\author[6,7]{\small Cristian Antonelli}
\author[1]{\small Leif K. Oxenløwe}
\author[2,3]{\small Alessandro Zavatta}

\affil[1]{\footnotesize CoE SPOC, DTU Fotonik, Technical University of Denmark, 2800 Kgs. Lyngby, DK}
\affil[2]{\footnotesize Istituto Nazionale di Ottica (CNR-INO), Largo E. Fermi 6, 50125 Florence, IT}
\affil[3]{\footnotesize LENS and Department of Physics and Astronomy, University of Florence, 50019 Sesto Fiorentino, IT}
\affil[4]{\footnotesize Department of Physics “Ettore Pancini", University of Naples “Federico II", Via Cinthia 21, 80126 Naples, IT}
\affil[5]{\footnotesize Optical Communications Laboratory, Sumitomo Electric Industries, Ltd., 244-8588 Yokohama, JP}
\affil[6]{\footnotesize Department of Physical and Chemical Sciences, University of L'Aquila, L'Aquila, IT}
\affil[7]{\footnotesize National Laboratory of Advanced Optical Fibers for Photonics (FIBERS), CNIT, L’Aquila, Italy}

\affil[*]{dabac@fotonik.dtu.dk}

\date{} 
\pagestyle{plain}
\setcounter{page}{1}
\twocolumn[ 
\begin{@twocolumnfalse}
\maketitle
     \vspace{-0.8cm}
\begin{abstract}
\normalsize
\vspace*{-1.0em}
\noindent
{Multicore fibers are expected to be a game-changer in the coming decades thanks to their intrinsic properties, allowing a larger transmission bandwidth and a lower footprint in optical communications. In addition, multicore fibers have recently been explored for quantum communication, attesting their uniqueness in transporting high-dimensional quantum states. However, investigations and experiments reported in literature have been carried out in research laboratories, typically making use of short fiber links in controlled environments. Thus, the possibility of using long distance multicore fibers for quantum applications is still to be proven.
We here characterize for the first time, in terms of phase stability, multiple strands of a 4-core multicore fiber installed underground in the city of L'Aquila, with an overall fiber length up to about 25 km. In this preliminary study, we investigate the possibility of using such an infrastructure to implement quantum-enhanced schemes, such as high-dimensional quantum key distribution, quantum-based environmental sensors, and more in general quantum communication protocols. }

\vspace{0.5cm}
\end{abstract}
  \end{@twocolumnfalse} ]

\begin{figure*}[t]
\centering
\includegraphics[width=0.9\textwidth]{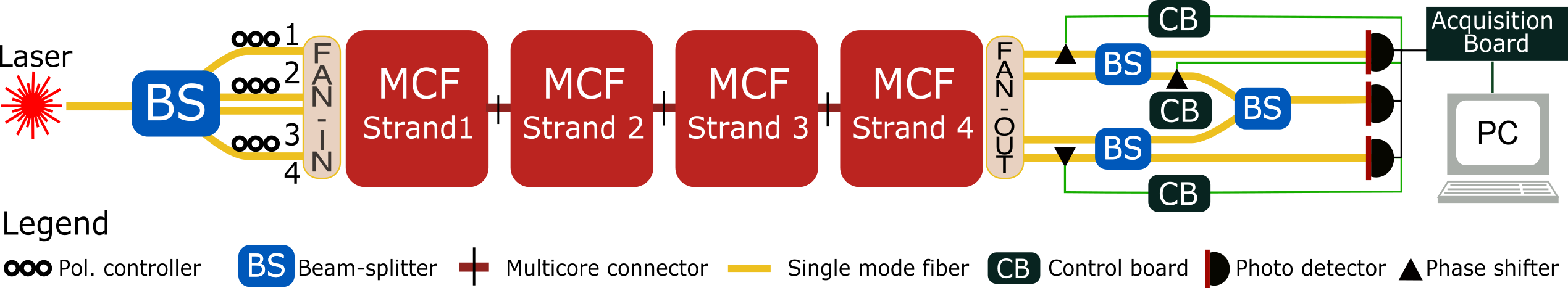}
\caption{{\bf Experimental setup.} A continuous wave laser at 1550 nm is equally divided into four paths through a 1x4 beam splitter (BS). The four single-mode fibers are then individually connected to the fan-in input of the multicore fiber. Thanks to the reconfigurability of the optical system, it is possible to set the number of subsequent MCF strands to be tested in the experiment (each strand is approximately 6.29 km of length). To connect the strands to each other, a multicore fiber connector is used. After propagation through the MCF, a fan-out device is used to divide the cores into four different single-mode fibers, that are finally combined with each other by three 2x2 beam splitters. By using three automatic control boards, each driving a phase shifter, based on the measurement output of the three detectors, we can monitor and individually control the relative phase of each core. %The light after is acquired by three photo-diode which are connected to a Nidaq acquisition board, controlled by a computer.}
}
\label{fig:1}
\end{figure*}

%microcontroller;

\section*{Introduction}
\vspace{-0.25cm}
%\nb{(Usiamo strands invece che spools. è più chiaro e è quello che usano in tutti i loro articoli. OK, lo cambio anche nelle figure)}{}
%\newline
Multicore fibers (MCFs), i.e., optical fibers with multiple cores within the same cladding, are expected to be a game-changer in the next generation of telecommunication infrastructures\cite{saitoh2016,amma2015high,hu2018single,hayashi2019field}. In fact, multicore fibers present multiple advantages over standard single-mode fibers, while maintaining, at the same time, a similar performance in terms of optical loss \cite{hayashi2011low,hayashi2017record}. Specifically, multicore fibers constitute a promising candidate for the implementation of space-division multiplexed transmission, and their standardization is taken into considered\cite{itu_std_2021}. They present a lower footprint, which is of high importance in the deployment process, e.g., in telecom data-centers where the space is limited, and they allow the use of a single amplifier for all the cores, reducing the number of resources\cite{yuan2018space}.
Furthermore, multicore fibers represent a perfect match with photonic integrated circuits for multiple-input-multiple-output applications, and are expected to be widely adopted in long-haul undersea connections\cite{nooruzzaman2017multi}. 
Recently, multicore fibers have also been tested for quantum communication\cite{dynes2016quantum,bacco2019,da2018record,ding2017,canas2017,bacco2017space,xavier2020quantum,da2019stable}. Thanks to their properties of low loss and small crosstalk between the different cores, these fibers, which are referred to as uncoupled-core MCFs, have been used for co-propagating quantum and classical signals in different cores\cite{dynes2016quantum}, or in the same core \cite{bacco2019,da2018record}, and also for transmitting high-dimensional quantum states\cite{ding2017,canas2017,bacco2017space,xavier2020quantum,da2019stable}. High-dimensional quantum states, thanks to their intrinsic properties, allow for a higher information capacity (useful in the case of a limited photon budget or in the regime of saturating single-photon detectors) and also exhibit higher robustness to the noise affecting the quantum communication (which is critical in real-world applications)\cite{cozzolino2019review}.
%problema nello stabilizzazione
However, the transmission of high-dimensional quantum states over multicore fibers requires phase stability between the different cores, since the quantum states are encoded in coherent superpositions of the cores of the fiber. In fact, although the improved phase stability of a single multicore fiber, compared to a bundle of single-core fibers, was already demonstrated over 2 km of a 7-core uncoupled fiber (in laboratory environment), the phase stability of longer deployed MCFs was never tested so far\cite{da2019stable,dalio2021,alarcon2021fewmode}.
%nessun test fatto su fibra deplyed in fase
In this work, we study the phase stability of a 4-core uncoupled multicore fiber with 25.16 km of length, deployed in the city of L'Aquila\cite{hayashi2019field}. Furthermore, we propose and demonstrate a scalable fully-automatic stabilization method which is potentially useful in transporting and manipulating the high-dimensional quantum states over a multicore fiber. %In addition, we suggest some broader applications for such a test-bed facility. 

\section*{Experimental setup}
\vspace{-0.25cm}

%\subsection*{Field-deployed multicore fiber testbed}
%As part of the INCIPIT \nb{(ref)}{} project, an optical testbed infrastructure was built, enabling researchers from around the world to perform experiments on multicore fibers deplyed under the city of L'Aquila.
In our experiment we have used an optical test-bed infrastructure built in 2019 in the city of L'Aquila\cite{hayashi2019field}. A single jelly-filled loose-tube cable, with an outer diameter of $6$ mm and a total length of $6.29$ km, has been deployed in a multi-service underground tunnel with both ends accessible from the same location. 
The cable accommodates three different kinds of multicore fibers  for a total of 18 strands: twelve are randomly-coupled 4-cores MCFs (RC-4CF), four are uncoupled 4-cores MCFs (UC-4CF) and two are uncoupled 8-cores MCFs (UC-8CF). The first two kinds (RC-4CF and UC-4CF) are optimized for the C-band window (1550 nm), while the latter (UC-8CF) is optimized for the O-band window (1310 nm). 
All the cores in each strand are accessible via an optical patch panel. More details are reported in the work by T. Hayashi and co-authors \cite{hayashi2019field}. In our experiment, we focused on the UC-4CF strands, which to date have been characterized only with respect to inter-core crosstalk \cite{hayashi2019field} and time skew \cite{Luis:20,Puttnam:20}. 
All the UC-4CF strands are terminated by spliced SC-connectorized MCF pigtails on both ends. This allows to concatenate multiple strands, up to a total MCF length of $25.16$ km, or to access each core individually by connecting the strand to a fan-in fan-out device, that splits the four cores into four different single-mode fibers. %\nb{XX}{}-connectorized.

\begin{figure}[t]
\centering
\includegraphics[width=0.5\textwidth]{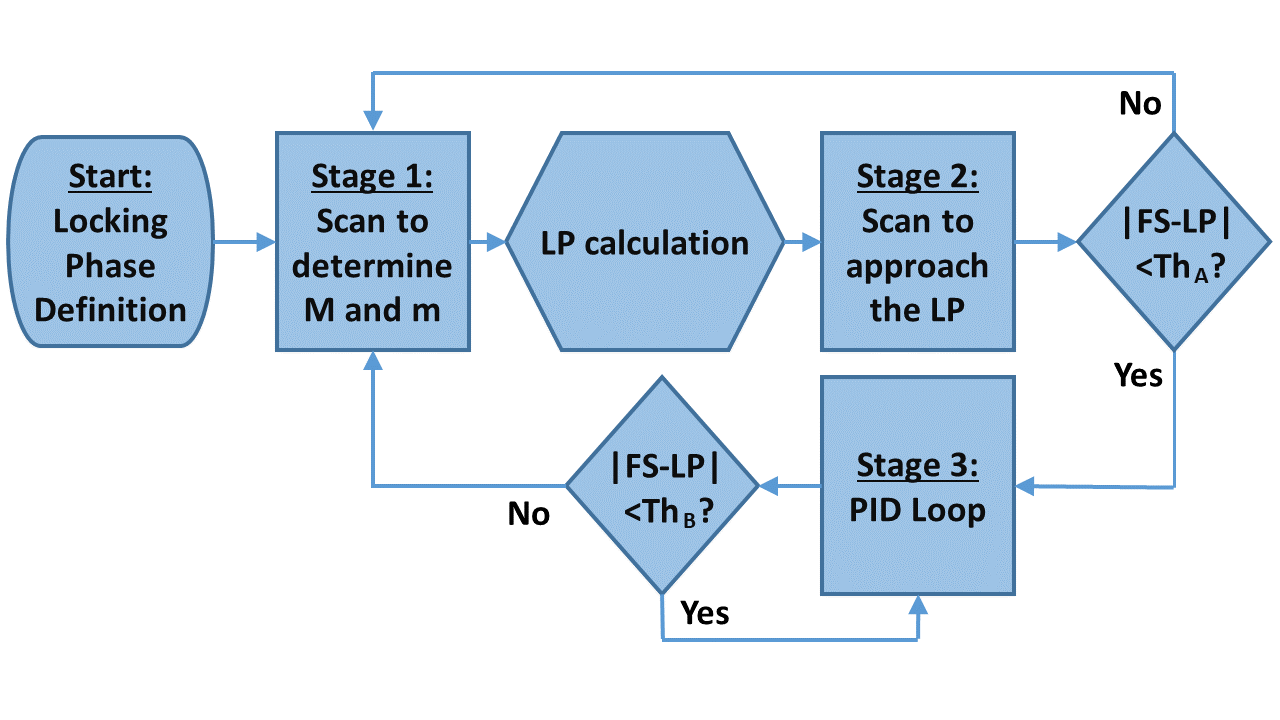}
\caption{\textbf{Flow chart of the PID controller}. This algorithm is used for controlling each phase-shifter in the three different interferometers.}
\label{fig:flawCh}
\end{figure}

%\subsection*{Measurement scheme}
In order to adopt the MCF for quantum applications, an efficient stabilization of the relative phase between all the different cores is required. Figure \ref{fig:1} shows the interferometric scheme we have implemented to test the phase stability of the UC-4CF. 
%deployed under the city of L'Aquila.
%%possibility of using a non-commercial phase lock loop (PLL) to efficiently stabilize 2-paths and 4-paths \nb{multicore fiber-based (???)}{} interferometers. 
The output of a continuous wave laser emitting at 1550 nm has been injected into the common port of a 1x4 fiber beam-splitter (BS). After the beam-splitter, we have inserted 3 fiber-based polarization controllers to correctly aligned the polarization inside the different cores of the fiber. The output ports have been connected to the four cores of a UC-4CF strand, through the fan-in device. The end of this strand can be connected directly to the last part of the interferometric setup, by using the fan-out kit, or to the input of another UC-4CF strand. In this way, we can adjust the overall interferometer length from $6.29$ km to $25.16$ km, by steps of $6.29$ km. The four cores of the last connected strand, after the fan-out kit, have been injected into the input ports of two 2x2 BSs. Thus, two independent 2-path interferometers between cores 1 \& 2 and cores 3 \& 4, respectively, have been realized. By monitoring one output port of each interferometer,  we can measure the relative phase fluctuations between cores 1 $\&$ 2 and cores 3 $\&$ 4, as shown in Figure \ref{fig:1}. Moreover, by combining the remaining output ports of the 2-path interferometers via a third 2x2 BS, we can access the relative phases between all of the cores. %The light at the outputs of the three interferometers was measured using a XXX photo-detector
%{The light in all the three interferometers was measured adopting a photo-detector} (insert specs). \nb{}{Subsequently, the electrical signal was divided in order to properly control the interferometers and to acquire the measurements. The measurement were} 
The three interferometer outputs were monitored with p-i-n photodiodes, resulting in three electrical signals that were collected with a {\it National Instruments} Data Acquisition Board (NIDaq) , connected to a computer (PC).\\
%The output of the photo-detectors was acquired using a NIDaq acquisition board connected to a PC.
In order to fully and individually control the relative phases between the four uncoupled cores, a Phase Lock Loop (PLL) has been integrated in each interferometer. 
%suppers these fluctuations, a Phase Lock Loop (PLL) has been integrated in each interferometer.  
The main component of each PLL is the control board (CB) composed by an ADuC7020 microcontroller unit (MCU), produced by Analog Devices. This device incorporates a 5-channels 12-bit analog-to-digital converter (ADC) and a 4-channels 12-bit digital-to-analog converter (DAC). Thanks to the high resolution of the analog channels, and to the MCU clockrate of 41.78 MHz, this microcontroller is suitable to implement a digital proportional-integral-derivative controller (PID), able to efficiently stabilize the MCF interferometers. In order to maximize the computational power available to stabilize each interferometer, we used a dedicated ADuC7020 for each PLL. The feedback signal (FS) used by the PLL is a portion of the output of the photo-detector monitoring the same interfeometer the PLL is controlling. An ADC channel, preceded by a low-pass filter with a bandwidth of 1 KHz, is used to acquire this signal.
%{ An ADC channel is used to monitor the feedback signal (FS) coming from the photo-diode placed at the output port of each interferometer.} 
The intensity of this signal can be related to the relative phase $\varphi$ between the cores that are involved in the 2-path interferometer:

\begin{equation}
    FS(\varphi)=\frac{M-m}{2} \cos(\varphi)+\frac{M+m}{2},
\label{eq:FS}
\end{equation}
where $M$ and $m$ are the maximum and the minimum, respectively, of the interference fringes. In order to actively control this phase, a fiber-based phase-shifter has been introduced in each arm of each interferometer, as shown in Figure \ref{fig:1}. A DAC channel is used to control, via an high-voltage driver, the phase shifter. The firmware installed on the microcontroller can be divided into two main blocks: one generates a triangular ramp at the DAC output to uniformly scan the interferometer relative phase, the other contains the code that implements the digital PID used for the phase stabilization. %The first block is essentially a counter that linearly increase and decrease the DAC voltage from 0V to 2.5V and vice-versa, the locking block is more complex. 
Its behaviour is described in the flowchart in Figure \ref{fig:flawCh}. 

The PID block is divided into three main stages. The starting point is a serial input from the user, that communicates to the microcontroller the locking phase, i.e., the phase where the interferometer should be locked. This command activates the first stage, where the voltage applied to the phase-shifter is scanned, in order to produce a phase variation slightly bigger than $2\pi$. Meanwhile, the FS is measured by the microcontroller ADC to find the maximum ($M$) and the minimum ($m$) of the interference fringes. By knowing these two parameters, it is possible to invert Equation \ref{eq:FS} to determine the value of the FS corresponding to the desired locking phase, called Locking Point (LP). The locking phase value, i.e. the LP, is a parameter which can be arbitrarily fixed by the user in the range from 0 to $2 \pi$.
During the second stage, the same phase range is scanned in order to approach the locking point, with a precision determined by the user-defined threshold $Th_A$. By measuring the FS signal during this scan, it is also possible to select the slope of the interference fringe used for locking, which allows us to stabilize the relative phase over the all range $[0, \ 2\pi]$. As soon as $|FS-LP|<Th_A$, the PID loop \cite{ROZSA1989115} starts operating (stage 3). Otherwise, if this condition can not be reached, the algorithm restarts back from the first stage. Once activated, the PID loop keeps acting as long as the condition $|FS-LP|<Th_B$ remains verified. The $Th_B$ threshold fixes the maximum tolerable difference between FS and the desired LP. If this condition is not verified, the PID loop stops and the algorithm starts back from the first stage, allowing for an automatic re-locking. 

\begin{figure}[t]
\centering
\includegraphics[width=0.5\textwidth]{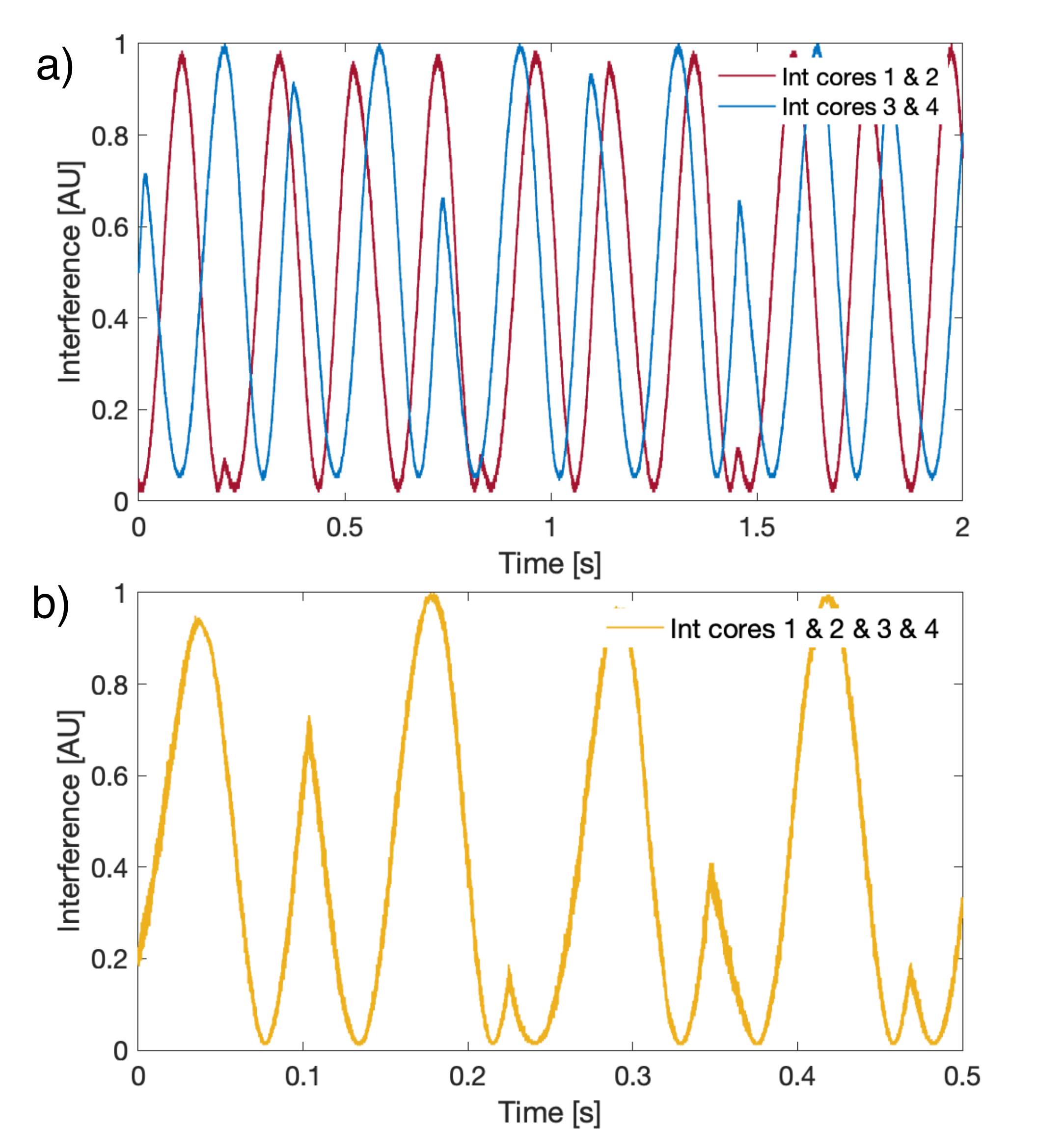}
\caption{{\bf Visibility fringes of the interferometers.} \textbf{a)} Interference signals as a function of time for the two independent interferometers between cores 1 $\&$ 2 and cores 3 $\&$ 4.  Measured visibility of $V_{1 \& 2} = 0.981 \pm 0.008$ and $V_{3 \& 4} = 0.945 \pm 0.011$. \textbf{b)} Interference signal from the three interferometers. In this measurement, the fist two independent interferometers are locked to a fixed position. Measured visibility of $V_{1 \& 2 \& 3 \& 4} = 0.989 \pm 0.004$.}
\label{fig:2}
\end{figure}

\begin{figure}[t]
\centering
\includegraphics[width=0.5\textwidth]{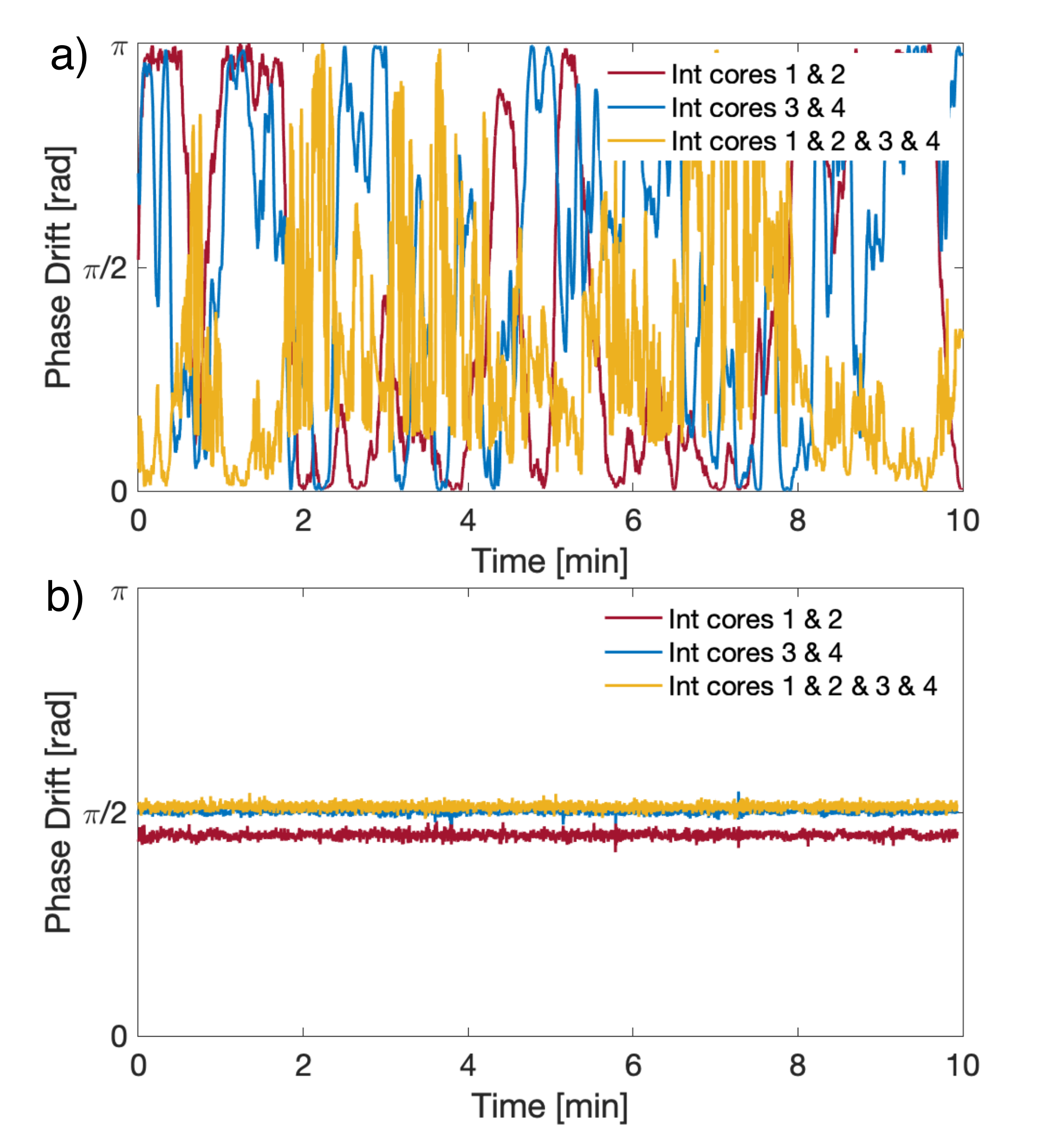}
\caption{{\bf Phase drifts of the non stabilized and stabilized multicore interferometers over 10 minutes acquisition.} \textbf{a)} Phase drift between two-cores and four-cores interferometers without active phase stabilization. \textbf{b)} Phase drift of a) but  with active phase stabilization loops. Different colours represent the three  different configurations (red and blue, two paths interferometers; yellow, four paths interferometer). }
\label{fig:3}
\end{figure}

%cambiare colori della figura

\section*{Results}
%%%%%%%%%%%%%%%%%%%%%%%%%%%%%%%%%%%%%%%%%%%%%%%%%%%%%%%%%
In order to phase-stabilize the cores of the multicore fiber, we built the three interferometers as described in the previous paragraph. The first parameter to be evaluated in these interferometers is the visibility, which is directly linked to the performance of a quantum or classical communication protocol, in terms of expected error rate. In Figure \vref{fig:2}a), the red (blue) curve shows the interference fringes obtained by driving, with a triangular shape, the fiber phase-shifter in the 2-path interferometer involving cores 1 \& 2 (3 \& 4) of the MCF. These two interferometers, as shown in the experimental setup in Figure \ref{fig:1}, are independent. We measured a visibility of $V_{1 \& 2} = 0.981 \pm 0.008$ and $V_{3 \& 4} = 0.945 \pm 0.011$. The third interferometer involves all of the four cores and, to evaluate its visibility, the two independent interferometers (involving cores 1 \& 2 and cores 3 \& 4 separately) must be locked to a specific phase, in order to balance the two powers entering in the third BS. Through the first two PLLs, we have locked the two independent interferometers to around 50\% of the fringe and, exploiting the third phase shifter, we generated a triangular ramp in the overall interferometer, as shown in Figure \ref{fig:2}b) with yellow color. The visibility value we measured is $V_{1 \& 2 \& 3 \& 4} = 0.989 \pm 0.004$. 

Subsequently to the visibility measurement, we investigated the possibility of stabilizing the four different cores of the MCF for a certain amount of time. The stabilization of the relative phase between the cores of a multicore fiber is an essential property for the reliable transmission of the quantum states, both in quantum key distribution applications but also in more advanced quantum network protocols. We report in Figure \ref{fig:3}a) the temporal drift of the relative phases between the cores, observed with the unlocked interferometers over a continuous and free-running acquisition of ten minutes.
Figure \ref{fig:3}b) shows the same acquisition in which we have turned on the three automatic PLL systems, to actively compensate the drifts. In order to stabilize all the different cores, we have first locked the two independent interferometers involving cores 1 \& 2 and cores 3 \& 4, and subsequently we have locked the overall interferometer. To be noted that, in case of fast and abrupt drifts in the fiber, i.e. when the locking position is suddenly lost, our phase lock loop system is able to automatically re-lock to the same position, by restarting the PLL algorithm from the first stage (see Figure \ref{fig:flawCh}). Another important point to be highlighted is that the polarization of the different cores was stable over 100 minutes of acquisition time.    

\begin{figure*}[ht!]
\centering
\includegraphics[width=1\textwidth]{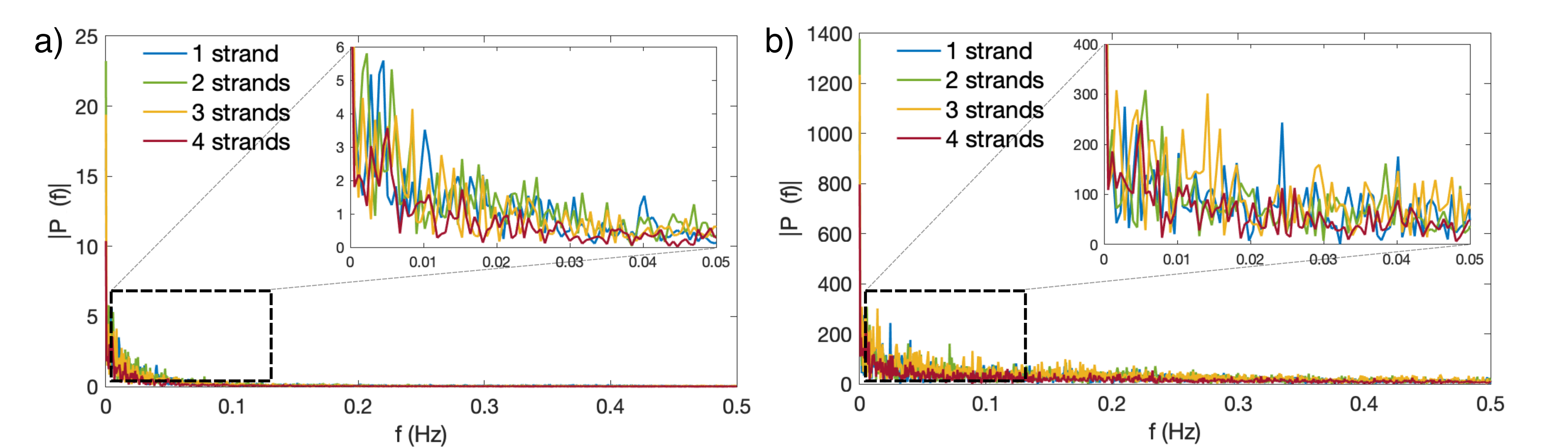}
\caption{{\bf Frequency analysis of the interference signals.} \textbf{a)} Two-cores intereferometers for the four different strands. \textbf{b)} Four-cores interferometer for the four different strands. Different colours represent different strands. Each measurement has been acquired for 30 minutes of non-locked system.}
\label{fig:4}
\end{figure*}

In addition, we further investigated the free-running acquisition in order to better characterize the signal phase fluctuations in the multicore fiber, for different fiber lengths. To this end, we made additional 30-minutes acquisitions, with 6 Hz sampling rate, of the cores interference signals for multiple strands. 
In Figure \ref{fig:4} we show the results of these measurements. In the left panel, we report the power spectrum
%density function 
of the interference signal as a function of frequency for one of the two-cores interferometers (1 \& 2). In the right panel we report the same measurement for the four-cores interferometer.

%togliere la misura a 3 inteferometri
%fare inset

\section*{Discussion}
Optical interferometers are the basic component for optical signal processing. More specifically, fiber-based interferometers are widely used for different applications, spanning from sensing, and optical communication to quantum physics and gravitational wave detections. 
Multipaths interferometers are used for manipulating quantum states and high-dimensional unitary operations, and in this work we demonstrated the possibility of stabilizing a long-distance multicore fiber through a simple and scalable setup. For example, we might imagine to use a wavelength multiplexing approach for transmitting quantum and classical light in the same fiber. One of these channels could be used for stabilizing the drift of the relative phase as demonstrated in \cite{dalio2021}
In the same direction, we have also demonstrated that the phase drift is not directly related to the length of the multicore fiber. In fact, by looking at Figure \ref{fig:4}, both configurations seem quite insensitive to the overall interferometric size. This behaviour can be explained by considering that most of the phase fluctuations comes from the fibers connecting the rack-mount optical patch panel to the rest of the experimental setup, located on a table as close to it as possible. In other words, we can assume that the most unstable portion of the interferometric apparatus is the fan-in and fan-out part \cite{Hu2020}.
%\nb{(ask Mecozzi if they have an explanation why 4-spools configurations seem more stable than shorter ones.)}{} 

The second evidence, from Figure \ref{fig:4}b), is that the 4-cores interferometer is more sensitive to phase fluctuations than the 2-path configurations. In fact, the spectral density function of of the 4-cores interferometer is about 2 order of magnitudes higher as compared to the 2-paths one. This fact, as already demonstrated in \cite{zhao2018robust,d1997arbitrary,gan2016spatial,schreiber2013invited,weihs1996all}, could be quite useful for sensing applications, both classical and quantum.

%lima Multiport BS 
To be noted that in our demonstration, we have used cascaded interferometers for analysing independently all the optical signals, but new devices have been recently introduced for multi-port beam splitter \cite{pereira2020universal}. These devices are properly designed for acting as interfaces between single-mode fibers and multicore fibers, and could increase the overall stability of the system.

Summarizing, we here presented a scalable and efficient method for stabilizing the phase drifts in multicore fiber. The presented method can in principle be applied to longer fiber distances and larger core counts, by using the same technology. Our demonstration paves the way towards future investigations and applications of multicore fibers in quantum communication.   

\section*{Acknowledgements}
\vspace{-0.25cm}

\section*{Funding}
\vspace{-0.25cm}
This work is supported by the Center of Excellence, SPOC - Silicon Photonics for Optical Communications (ref DNRF123), by the EraNET Cofund Initiatives QuantERA within the European Union’s Horizon 2020 research and innovation program grant agreement No.731473 (project SQUARE), by the NATO Science for Peace and Security program  under grant G5485, by the Italian Government through Project INCIPICT and Project PRIN 2017 FIRST, and from the European Union through OpenQKD-project MuQuAKE (project number: 857156).

%\section*{Abbreviations}
%QKD Quantum key distribution; BB84 Bennett and Brassard 1984 QKD protocol; 2D two-dimensional; 4D four-dimensional; IM Intensity Modulator; CW continuous wave; PM Phase Modulator; BS beam splitter; PBS polarizing beam splitter; VOA variable optical attenuator; FPGA Field Programmable Gate Array; PRBS pseudo random binary sequence; SPAD Single Photon Avalanche Detectors; QBER Quantum Bit Error Rate; SKR Secret Key Rate.

%\section*{Availability of data and materials}
%Not applicable. For all requests relating to the paper, please contact the first author.

%\section*{Competing interests}
%The authors declare that they have no competing interests.

%\section*{Authors’ contributions}
%D.B. and I.V. conceived the experiment. I.V., B.D.L., D.C., and D.B. carried out the experimental work. D.R. and B.D.L. carried out the theoretical analysis on the protocols. All authors contributed to the writing of the manuscript.

%\newpage
%\bibliography{mybib}{}

\begin{thebibliography}{30}

\bibitem{saitoh2016}
K.~Saitoh and S.~Matsuo, ``Multicore fiber technology,'' {\em Journal of
  Lightwave Technology}, vol.~34, no.~1, pp.~55--66, 2016.

\bibitem{amma2015high}
Y.~Amma, Y.~Sasaki, K.~Takenaga, S.~Matsuo, J.~Tu, K.~Saitoh, M.~Koshiba,
  T.~Morioka, and Y.~Miyamoto, ``High-density multicore fiber with
  heterogeneous core arrangement,'' in {\em 2015 Optical Fiber Communications
  Conference and Exhibition (OFC)}, pp.~1--3, IEEE, 2015.

\bibitem{hu2018single}
H.~Hu, F.~Da~Ros, M.~Pu, F.~Ye, K.~Ingerslev, E.~P. da~Silva, M.~Nooruzzaman,
  Y.~Amma, Y.~Sasaki, T.~Mizuno, {\em et~al.}, ``Single-source chip-based
  frequency comb enabling extreme parallel data transmission,'' {\em Nature
  Photonics}, vol.~12, no.~8, pp.~469--473, 2018.

\bibitem{hayashi2019field}
T.~Hayashi, T.~Nagashima, T.~Nakanishi, T.~Morishima, R.~Kawawada, A.~Mecozzi,
  and C.~Antonelli, ``Field-deployed multi-core fiber testbed,'' in {\em 2019
  24th OptoElectronics and Communications Conference (OECC) and 2019
  International Conference on Photonics in Switching and Computing (PSC)},
  pp.~1--3, IEEE, 2019.

\bibitem{hayashi2011low}
T.~Hayashi, T.~Taru, O.~Shimakawa, T.~Sasaki, and E.~Sasaoka, ``Low-crosstalk
  and low-loss multi-core fiber utilizing fiber bend,'' in {\em Optical Fiber
  Communication Conference}, p.~OWJ3, Optical Society of America, 2011.

\bibitem{hayashi2017record}
T.~Hayashi, Y.~Tamura, T.~Hasegawa, and T.~Taru, ``Record-low spatial mode
  dispersion and ultra-low loss coupled multi-core fiber for ultra-long-haul
  transmission,'' {\em Journal of Lightwave Technology}, vol.~35, no.~3,
  pp.~450--457, 2017.

\bibitem{itu_std_2021}
.~I.-T. S. W.~Q. meeting, ``Standardization of optical fibre cable and
  components for sdm technology,'' 2020.

\bibitem{yuan2018space}
H.~Yuan, M.~Furdek, A.~Muhammad, A.~Saljoghei, L.~Wosinska, and G.~Zervas,
  ``Space-division multiplexing in data center networks: on multi-core fiber
  solutions and crosstalk-suppressed resource allocation,'' {\em Journal of
  Optical Communications and Networking}, vol.~10, no.~4, pp.~272--288, 2018.

\bibitem{nooruzzaman2017multi}
M.~Nooruzzaman and T.~Morioka, ``Multi-core fibers in submarine networks for
  high-capacity undersea transmission systems,'' in {\em 2017 Optical Fiber
  Communications Conference and Exhibition (OFC)}, pp.~1--3, IEEE, 2017.

\bibitem{dynes2016quantum}
J.~Dynes, S.~Kindness, S.-B. Tam, A.~Plews, A.~Sharpe, M.~Lucamarini,
  B.~Fr{\"o}hlich, Z.~Yuan, R.~Penty, and A.~Shields, ``Quantum key
  distribution over multicore fiber,'' {\em Optics express}, vol.~24, no.~8,
  pp.~8081--8087, 2016.

\bibitem{bacco2019}
D.~Bacco, B.~Da~Lio, D.~Cozzolino, F.~Da~Ros, X.~Guo, Y.~Ding, Y.~Sasaki,
  K.~Aikawa, S.~Miki, H.~Terai, {\em et~al.}, ``Boosting the secret key rate in
  a shared quantum and classical fibre communication system,'' {\em
  Communications Physics}, vol.~2, no.~1, pp.~1--8, 2019.

\bibitem{da2018record}
B.~Da~Lio, D.~Bacco, D.~Cozzolino, F.~Da~Ros, X.~Guo, Y.~Ding, Y.~Sasaki,
  K.~Aikawa, S.~Miki, H.~Terai, {\em et~al.}, ``Record-high secret key rate for
  joint classical and quantum transmission over a 37-core fiber,'' in {\em 2018
  IEEE photonics conference (IPC)}, pp.~1--2, IEEE, 2018.

\bibitem{ding2017}
Y.~Ding, D.~Bacco, K.~Dalgaard, X.~Cai, X.~Zhou, K.~Rottwitt, and L.~K.
  Oxenl{\o}we, ``High-dimensional quantum key distribution based on multicore
  fiber using silicon photonic integrated circuits,'' {\em npj Quantum
  Information}, vol.~3, no.~1, p.~25, 2017.

\bibitem{canas2017}
G.~Ca{\~n}as, N.~Vera, J.~Cari{\~n}e, P.~Gonz{\'a}lez, J.~Cardenas,
  P.~Connolly, A.~Przysiezna, E.~G{\'o}mez, M.~Figueroa, G.~Vallone, {\em
  et~al.}, ``High-dimensional decoy-state quantum key distribution over
  multicore telecommunication fibers,'' {\em Physical Review A}, vol.~96,
  no.~2, p.~022317, 2017.

\bibitem{bacco2017space}
D.~Bacco, Y.~Ding, K.~Dalgaard, K.~Rottwitt, and L.~K. Oxenl{\o}we, ``Space
  division multiplexing chip-to-chip quantum key distribution,'' {\em
  Scientific reports}, vol.~7, no.~1, pp.~1--7, 2017.

\bibitem{xavier2020quantum}
G.~B. Xavier and G.~Lima, ``Quantum information processing with space-division
  multiplexing optical fibres,'' {\em Communications Physics}, vol.~3, no.~1,
  pp.~1--11, 2020.

\bibitem{da2019stable}
B.~Da~Lio, D.~Bacco, D.~Cozzolino, N.~Biagi, T.~N. Arge, E.~Larsen,
  K.~Rottwitt, Y.~Ding, A.~Zavatta, and L.~K. Oxenl{\o}we, ``Stable
  transmission of high-dimensional quantum states over a 2-km multicore
  fiber,'' {\em IEEE Journal of Selected Topics in Quantum Electronics},
  vol.~26, no.~4, pp.~1--8, 2019.

\bibitem{cozzolino2019review}
D.~Cozzolino, B.~Da~Lio, D.~Bacco, and L.~K. Oxenl{\o}we, ``High-dimensional
  quantum communication: Benefits, progress, and future challenges,'' {\em
  Advanced Quantum Technologies}, p.~1900038, 2019.

\bibitem{dalio2021}
B.~Da~Lio, D.~Cozzolino, B.~Nicola, Y.~Ding, K.~Rottwitt, A.~Zavatta, D.~Bacco,
  and L.~K. Oxenløwe, ``Path-encoded high-dimensional quantum communication
  over a 2 km multicore fiber,'' {\em arXiv:2103.05992}, 2021.

\bibitem{alarcon2021fewmode}
A.~Alarcón, J.~Argillander, G.~Lima, and G.~B. Xavier, ``Few-mode fibre
  technology fine-tunes losses of quantum communication systems,'' 2021.

\bibitem{Luis:20}
R.~S. Lu\'{i}s, B.~J. Puttnam, G.~Rademacher, A.~Marotta, C.~Antonelli,
  F.~Graziosi, A.~Mecozzi, T.~Hayashi, T.~Nakanishi, S.~Shinada, Y.~Awaji,
  H.~Furukawa, and N.~Wada, ``Evaluation of dynamic skew on spooled and
  deployed multicore fibers using o-band signals,'' in {\em Optical Fiber
  Communication Conference (OFC) 2020}, p.~T4J.4, Optical Society of America,
  2020.

\bibitem{Puttnam:20}
B.~J. Puttnam, R.~S. Luis, G.~Rademacher, A.~Marotta, C.~Antonelli, A.~Mecozzi,
  F.~Graziosi, T.~Hayashi, T.~Nakanishi, Y.~Awaji, H.~Furukawa, and N.~Wada,
  ``Dynamic skew measurements in a deployed 4-core fiber,'' in {\em Conference
  on Lasers and Electro-Optics}, p.~STu4R.1, Optical Society of America, 2020.

\bibitem{ROZSA1989115}
L.~Rozsa, ``Design and implementation of practical digital pid controllers,''
  {\em IFAC Proceedings Volumes}, vol.~22, no.~18, pp.~115--121, 1989.

\bibitem{Hu2020}
X.-M. Hu, W.-B. Xing, B.-H. Liu, D.-Y. He, H.~Cao, Y.~Guo, C.~Zhang, H.~Zhang,
  Y.-F. Huang, C.-F. Li, and G.-C. Guo, ``Efficient distribution of
  high-dimensional entanglement through 11 km fiber,'' {\em Optica}, vol.~7,
  pp.~738--743, Jul 2020.

\bibitem{zhao2018robust}
Z.~Zhao, Z.~Liu, M.~Tang, S.~Fu, L.~Wang, N.~Guo, C.~Jin, H.-Y. Tam, and C.~Lu,
  ``Robust in-fiber spatial interferometer using multicore fiber for vibration
  detection,'' {\em Optics express}, vol.~26, no.~23, pp.~29629--29637, 2018.

\bibitem{d1997arbitrary}
G.~M. D'Ariano and M.~G. Paris, ``Arbitrary precision in multipath
  interferometry,'' {\em Physical Review A}, vol.~55, no.~3, p.~2267, 1997.

\bibitem{gan2016spatial}
L.~Gan, R.~Wang, D.~Liu, L.~Duan, S.~Liu, S.~Fu, B.~Li, Z.~Feng, H.~Wei,
  W.~Tong, {\em et~al.}, ``Spatial-division multiplexed mach--zehnder
  interferometers in heterogeneous multicore fiber for multiparameter
  measurement,'' {\em IEEE Photonics Journal}, vol.~8, no.~1, pp.~1--8, 2016.

\bibitem{schreiber2013invited}
K.~U. Schreiber and J.-P.~R. Wells, ``Invited review article: Large ring lasers
  for rotation sensing,'' {\em Review of Scientific Instruments}, vol.~84,
  no.~4, p.~041101, 2013.

\bibitem{weihs1996all}
G.~Weihs, M.~Reck, H.~Weinfurter, and A.~Zeilinger, ``All-fiber three-path
  mach--zehnder interferometer,'' {\em Optics letters}, vol.~21, no.~4,
  pp.~302--304, 1996.

\bibitem{pereira2020universal}
L.~Pereira, A.~Rojas, G.~Ca{\~n}as, G.~Lima, A.~Delgado, and A.~Cabello,
  ``Universal multi-port interferometers with minimal optical depth,'' {\em
  arXiv preprint arXiv:2002.01371}, 2020.

\end{thebibliography}
%\bibliographystyle{plain}
%\bibliographystyle{unsrt}
%\bibliographystyle{ieeetr}

\end{document}